\documentclass{midl} 

\usepackage{mwe} 
\usepackage{graphicx}
\usepackage{booktabs}
\usepackage{enumitem}
\jmlrvolume{-- Under Review}
\jmlryear{2024}
\jmlrworkshop{Full Paper -- MIDL 2024 submission}
\usepackage[subtle]{savetrees}
\usepackage{tabularray}
\usepackage{caption}

\midlauthor{\Name{Mohammed Adnan\midljointauthortext{Work done at Roche.}\nametag{$^{1}$}} \Email{adnan.ahmad@ucalgary.ca}\\
\addr $^{1}$ University of Calgary, Canada \AND
\Name{Qinle Ba\midljointauthortext{Corresponding author.}\nametag{$^{2}$}} \Email{qinle.ba@roche.com}\\
\Name{Nazim Shaikh\nametag{$^{2}$}} \Email{nazim.shaikh@roche.com}\\
\Name{Shivam Kalra\nametag{$^{2}$}} \Email{shivam.kalra@roche.com}\\
\Name{Satarupa Mukherjee\nametag{$^{2}$}} \Email{satarupa.mukherjee@roche.com}\\
\Name{Auranuch Lorsakul\nametag{$^{2}$}} \Email{auranuch.lorsakul@roche.com}\\
\addr $^{2}$ Roche Sequencing Solutions, Santa Clara, CA, USA 
}

\begin{document}
\title[Structure Model Pruning for DP]{Structured Model Pruning for Efficient Inference in Computational Pathology}

\maketitle

\begin{abstract}

\noindent Recent years have seen significant efforts to adopt Artificial Intelligence (AI) in healthcare for various use cases, from computer-aided diagnosis to ICU triage. However, the size of AI models has been rapidly growing due to scaling laws and the success of foundational models, which poses an increasing challenge to leverage advanced models in practical applications. It is thus imperative to develop efficient models, especially for deploying AI solutions under resource-constrains or with time sensitivity. One potential solution is to perform model compression, a set of techniques that remove less important model components or reduce parameter precision, to reduce model computation demand. In this work, we demonstrate that model pruning, as a model compression technique, can effectively reduce inference cost for computational and digital pathology based analysis with a negligible loss of analysis performance. To this end, we develop a methodology for pruning the widely used U-Net-style architectures in biomedical imaging, with which we evaluate multiple pruning heuristics on nuclei instance segmentation and classification, and empirically demonstrate that pruning can compress models by at least 70\% with a negligible drop in performance.
\end{abstract}

\section{Introduction}

With the recent trend to increase the capacity and thus performance of deep learning models in biomedical imaging~\cite{hörst2023cellvit, van2024exploring, deng2023segment}, it is increasingly challenging to fully leverage these effective yet computationally expensive models in real-world applications like clinical settings. Training advanced models and running inference at scale require dedicated computing resources, especially high-end GPUs. However, globally, the available GPU resources are unevenly distributed and especially rare in underdeveloped regions (https://www.top500.org/). In addition, the increasing computing needs require regular upgrade of IT infrastructure, imposing hurdles for AI adoption in healthcare institutions~\cite{zhang2022shifting}. High computing demand is especially challenging for digital and computational pathology (referred to as "DP") based applications: due to the large size of each whole-slide image (mega pixels), analysis is typically performed by first dividing an image into thousands of small image patches and then processing each patch with a model(s), followed by assembling patch-wise results into whole-side readouts~\cite{janowczyk2016deep}. As such, model inference can be time-consuming for a single image, let alone the large volumes of images in large medical centers~\cite{ardon2023digital}.

\indent{Model compression techniques reported in computer vision have shown remarkable success for reducing computation costs, which we classify into two broad categories: (1) techniques that require a predefined target architectures, which are more efficient than the original models, such as knowledge distillation~\cite{Hinton2015-wp} and neural architecture search (NAS)~\cite{ren2021comprehensive, baymurzina2022review}; (2) techniques that do not require such predefined architectures, such as model pruning~\cite{Cheng2023-aj}, quantization~\cite{Gholami2021-jn}. Many of these techniques are complementary and thus multiple techniques can be combined to achieve the best compression performance. For example, KD or NAS can be performed and followed by pruning and then by quantization. In this study, we investigate pruning because of its versatility. (1) Unlike KD or NAS, which requires design/selection of smaller network from larger network, pruning does not rely on such network dependency, and thus flexibly applicable without manual architectural modifications after KD or NAS. (2) Unlike quantization, which reduces parameter precision but keeps model architectures, pruning provides an opportunity to shrink architectures. (3) Pruning  achieves higher compression rates than typical KD methods~\cite{
javed2023knowledge, cho2019efficacy} (See Appendix~\ref{sec:appendix-KD} for KD in DP) and thus applying pruning on top of KD may further improve compression performance.}
In DP, \cite{CHOUDHARY2021104432} first studied \textit{structured filter pruning} for breast cancer classification, but they only assessed one pruning approach, L1-norm pruner.~\cite{10.1007/978-3-031-17721-7_12} applied \textit{unstructured magnitude pruning} for nuclei instance segmentation, 
which nominally "eliminate" a proportion of weights, but all the weight matrices still go through forward and backward passes and thus did not reduce the actual computation. To the best of our knowledge, no earlier works have systematically evaluated model pruning in DP for reducing computation costs. Moreover, no studies have investigated how to effectively prune the widely used U-Net style encoder-decoder architectures~\cite{ronneberger2015unet} in biomedical imaging. Such architectures can be challenging to prune: because of the shortcut connections between encoder and decoder layers and, between residual blocks in residual U-Nets \cite{alom2019recurrent}, pruning one layer triggers the necessity to manipulate the shortcut connected layers, potentially impacting many other layers. In this study, we propose a framework to handle such complex scenarios. Our contributions are as follows:
\begin{enumerate}[nosep]
\item We assess structured filter pruning with  recently proposed heuristics and strategies for compressing deep models for DP based analysis across two vision tasks, (1) nuclei instance segmentation and classification and (2) tile-level tissue classification.\item We propose a pruning approach for U-Net style architectures, a prevalent model design in biomedical imaging with plain convolution network or residual networks as the encoder.\item Using the proposed method, we compare different pruning heuristics on state-of-the-art models and show that pruning can effectively compress model size and reduce the latency of model inference with no or a minor decrease in model performance. 
\end{enumerate}
\section{Background}
\paragraph{Model Compression.}
Numerous strategies have been explored in the literature for model compression and four major categories include pruning, knowledge distillation (KD), quantization and neural 
architecture search (NAS). Pruning refers to removing relatively ‘less important’ model components based on 
certain heuristics to measure importance~\cite{blalock2020state}. Quantization refers to adopting less precise byte representation for model parameters and thus requiring less memory footprint, like reducing weight/gradient data type from \emph{float32} to \emph{int8}.
KD~\cite{Gou_2021} involves a larger model (teacher) and a smaller model (student). The teacher model is 
trained first and later used to guide the training of the smaller student model via distillation loss. NAS~\cite{ren2021comprehensive, baymurzina2022review} aims at automatically finding an efficient model architecture by searching through a design space. Many of these techniques are complementary and can be combined, among which is pruning. Here we focus on model pruning~\cite{blalock2020state}, which are further categorised 
into unstructured, semi-structured (See notes in Appendix~\ref{sec:appendix-semistruct}) and structured pruning, based on the sparsity patterns that a model pruning strategy targets. 

\paragraph{Unstructured Pruning.}
Determine the importance of each weight parameter and prune the relatively unimportant ones by replacing them with zero values. Unstructured pruning is straightforward to implement and usually does not harm model performance. However, current hardware is not optimized for unstructured 
sparse matrices and thus inference speed cannot be achieved without customized software/hardware systems. Therefore, we do not focus on such pruning methods.
\paragraph{Structured Pruning.} 
Structured pruning follows a fixed pattern. For example, an 
entire filter of a convolution layer or an entire layer can be removed. Contrary to unstructured pruning, since the entire layer~\cite{Ding2021-av,Fu2022-xz} or filter~\cite{Li2016-vt, Liu2017-zg} is removed, model speedup is achieved. Various pruning heuristics have been proposed. Here, we focus on the highly cited L1/L2 pruner \cite{Li2016-vt} and network slimmer \cite{Liu2017-zg}.

\paragraph{Pruning U-Net like Model Architectures.}
Prior work focused on classification models ~\cite{Liu2017-zg} for vision tasks. However, U-Net-based architectures are widely used in biomedical imaging due to their optimized design to encode information at multiple scales. Pruning U-Net style architecture is challenging due to shortcut connections between the encoder and decoder. In this work, we address this challenge by employing our pruning approach to the widely used architecture for nuclei segmentation and classification, HoverNet~\cite{Gamper2020-rc}.

\paragraph{HoverNet.}\cite{Gamper2020-rc} proposed HoverNet for nuclei instance segmentation and classification, which leverages the instance-aware information encoded by the vertical and horizontal distances
of nuclear pixels to their centres of mass. The corresponding distance maps are used to separate clustered nuclei, enabling accurate instance segmentation. For each segmented instance, the network predicts its nucleus type via a dedicated decoder branch. HoverNet has three branches, predicting nuclei segmentation, horizontal/vertical distance and nuclei classification.

\section{Methodology}

\begin{figure}[!hb]
    \centering
    \captionsetup{justification=raggedright,singlelinecheck=false}
    \includegraphics[width = 1.0\linewidth]{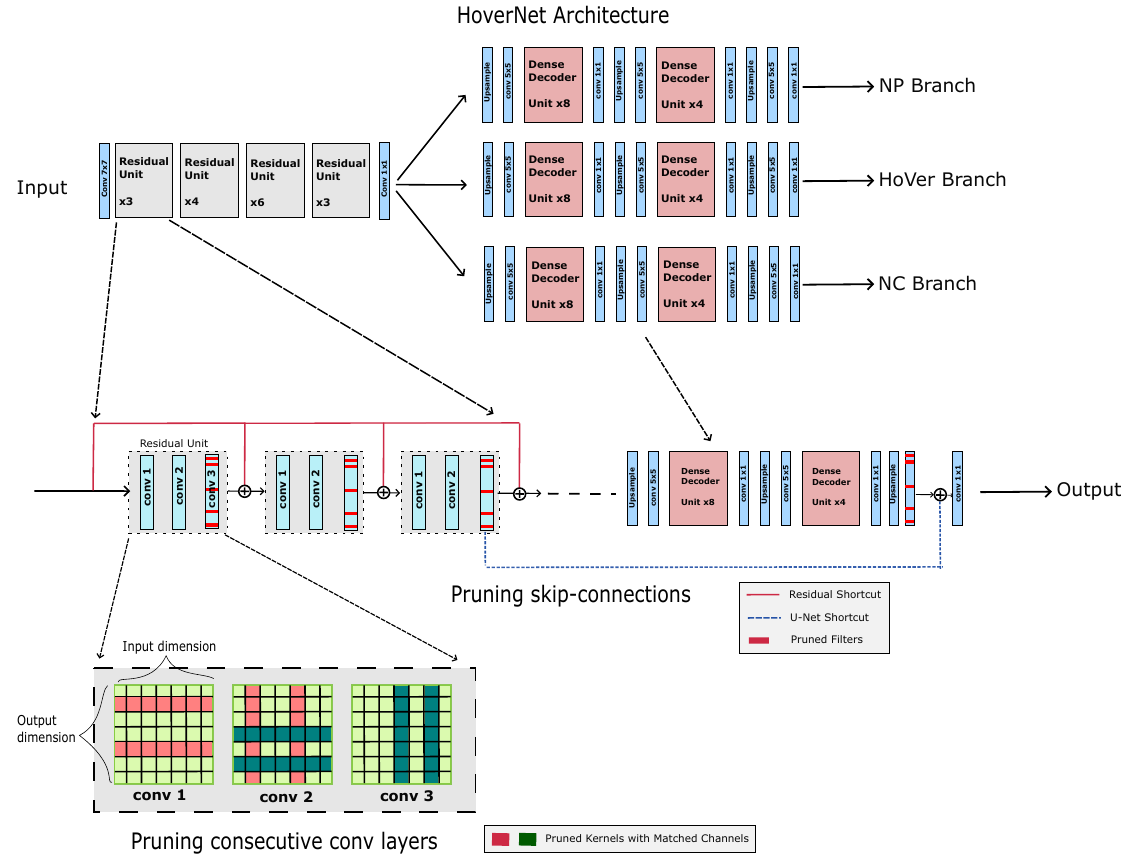}
    \caption{Overview of HoverNet and the proposed model pruning schema. 
    \underline{Top}: HoverNet design with three identical decoder branches. \underline{Middle}: dependencies between the last convolution (conv) layer of each residual block and the skip-connected decoder layer. Horizontal red bars denote pruned conv filters (output dimension) that are pruned with the matching channel indices (vertical bars) in interdependent layers. \underline{Bottom}: a 2D view of pruning consecutive conv layers. Here, each grid represents a 2D kernel (e.g. 3x3). The rows denote the output dimension (number of filters) and the columns denote the input dimension, which is equal to the number of feature maps of layer input. Pruning of conv filters from one layer (i.e red filters in first block) requires removing the kernels from the input dimension in the consecutive conv layer (i.e equivalent red filters in second block). Similarly, green filters illustrate pruning of the next two conv layers. }  
    
    
    \label{fig:enter-label}
\end{figure}

\paragraph{Pruning Heuristics and Strategies.}
Pruning heuristics refer to any approaches and/or metrics that determine the relative importance of target model components subjected to model pruning. By "pruning strategies" we refer to additional pruning approaches applied on top of heuristics.
\begin{enumerate}[nosep]
\item \underline{L1 Pruner}: ~\cite{Li2016-vt} removes the less important filters in a convolution layer based on the L1-norm of each filter, optionally followed by fine tuning to recover performance.

\item \underline{L2 Pruner}: Similar to L1 pruner, except that it uses L2-norm as \textit{pruning heuristic} to select less important filters to remove. \cite{Li2016-vt} found L1 and L2 norms to behave similarly.

\item \underline{Network Slimmer}: Proposed by \cite{Liu2017-zg}, applies L1-regularization on the scaling factors of batch normalization layers~\cite{Ioffe2015-fc} during baseline model training to induce sparsity of the scaling factors. Such a sparsity inducing regularization pushes less important scaling factors corresponding to the less important feature maps close to zero. It thus selects filters with smaller scaling factors after training as a \textit{pruning heuristic}. \textit{It is thus important to note that to the baseline for network slimmer is trained differently than that used for L1-norm and L2-norm pruner}.

\item \underline{Iterative Magnitude Pruning (IMP)}: A \textit{pruning strategy}~\cite{Frankle2019-pw} to stabilize the originally proposed Lottery Ticket Hypothesis~\cite{Frankle2018-kj}. IMP prunes a given network in multiple iterations instead of pruning at once. 
\end{enumerate}



\paragraph{Pruning HoverNet Model.}
HoverNet has a modified pre-activation ResNet50~\cite{He2015-ko} as encoder and a U-Net style design for connecting the encoder and decoders. Two types of shortcut connection in HoverNet include \underline{(1)} residual connections between the layers within each residual block in the 
encoder and \underline{(2)} skip connections connecting outputs of the encoder 
layers to decoder layers. The main challenge in pruning a U-Net style architecture is to ensure the dimensions of these layers with shortcut connection match after pruning. We thus propose the following approach for pruning. Since the 'residual' feature maps are presumably more important than the convolved ones within a residual block, we prioritize the former when applying heuristics \cite{Li2016-vt}.

\begin{enumerate}[nosep]
\item \underline{Handling Shortcut Connections between Encoder and Decoder Layers}: Use identical indices for pruning interconnected layers to maintain filter dimension compatibility post-pruning. This approach ensures that any interconnected layers are pruned in unison, preserving the structural integrity of the shortcut connections.
\item \underline{Pruning the Last Convolution Layers (Conv3) in Residual Blocks}: There are three convolution layers in each residual block in ResNet50 style encoders. The last convolution layer among the three, referred to as \emph{conv3} layer, is 
connected to the \emph{conv3} layer of the subsequent residual block due to the identity mapping or 1x1 convolution of residual connection mechanism. Last \emph{conv3} of each residual block is also skip connected to the decoder layers.
To ensure that the filter dimensions match for concatenation/addition in the shortcut 
connections, we prune \emph{conv3} from the encoder blocks and the corresponding skip connected layer in the decoder with the same filter indexing as depicted in second panel of~\autoref{fig:enter-label}.

\item \underline{Incorporating Non-Uniform Sparsity}: We found that pruning \emph{conv3} is challenging as pruning one \emph{conv3} layer triggers 
the pruning of \emph{conv3} layers in interconnected residual blocks and the skip-connected decoder layers. 
This may lead to over-pruning of important filters, as the actual filter importance of these interconnected and skip-connected layers are not assessed but rather follows the heuristic applied to the aforementioned \emph{conv3} layers. As such, we assessed non-uniform sparsity to limit the maximum sparsity level of the interconnected layers, aiming for flexible pruning rates across different layers of the network.
\end{enumerate}

\paragraph{Pruning Image Classification Models.}
Pruning of classification encoders are identical to the pruning of encoder part of an U-Net like encoder-decoder based model for dense prediction except that there's no skip connection to decoders. See Appendix \ref{sec:appendix-prune-resnet} for a detailed example.

\begin{enumerate}[nosep]
    \item \underline{Independent Pruning within Residual Block}: 
    Prune layers within each residual block or convolution block independently. In this option, only certain layers within each residual/convolution block that do not impact the subsequent layers are pruned, such as the first two convolution layers in each residual block of ResNet50. When pruning, it's crucial to consider the output dimension of the first convolution layer to prune, because it directly influences the input dimension of the following layer as depicted in third panel of \autoref{fig:enter-label}. 
With this pruning setting, one can choose the same or different pruning ratios for each block of layers (residual block, or a block composed of multiple convolution layers followed by other types of layers like batch-normalization layers), since the pruning of each block is independent. 

\item \underline{Pruning Interconnected Layers across Residual Blocks}: 
Pruning the last convolution layer in a residual block alters the output channel/filter dimension. As such, for the subsequent block, one must match the input feature maps' spatial and channel dimensions due to the addition operation connecting them with identity mapping. To maintain such compatibility, we ensure the pruning ratio and indexing of interconnected convolution layers to be consistent.
The channel number for the preceding batch-normalization layer should also be adjusted in line with the pruning performed on its subsequent convolution layer.  

\end{enumerate}

\section{Experiments and Results}
\subsection{Nuclei Instance Segmentation and Classification}
\textbf{Dataset and Implementation.} PanNuke dataset~\cite{Kingma2014-uz} contains image patches of 256x256 pixels selected from 20x or 40x H\&E stained whole-slide images of 19 tumor tissues with 189,744 segmented nucleus instances and 5 clinically important classes. We followed a 3-fold cross-validation~\cite{Kingma2014-uz} and trained HoverNet~\cite{Graham2018-oo} following the implementation, general training strategies and an pre-activation ResNet-50 based~\cite{he2016identity} encoder pre-trained on ImageNet~\cite{deng2009imagenet} provided by the HoverNet authors. Specifically, Adam optimizer was adopted with a learning rate of 0.0001 and weight decay of 0.0001. HoverNet training details and trained weights from three folds for PanNuke were not provided by the authors~\cite{Kingma2014-uz}, but, without applying an extensive list of tweaks, our model achieved a mean Panoptic Quality (mPQ) of 0.344 over all 5 tissue types (0.397 as reported by the authors). We assessed the following pruning strategies. \underline{(1)} One-shot uniform pruning with L1-norm, L2-norm and network slimmer as filter importance heuristics for pruning 10\% to 90\% of filters in target layers. \underline{(2)} One-shot non-uniform pruning with the L2-norm heuristic: the same setting as (1), but keeping an upper limit of 40\% filter pruning for interdependent layers across the residual blocks and across the encoder and decoder. \underline{(3)} Iterative pruning by removing a small percentage of filters at each pruning step: 5\% filter pruning for 19 rounds (5\% to 95\%). Each pruning step was followed by fine-tuning of 10 epochs with the same training settings. Model performance is evaluated with mean Panoptic Quality (mPQ), mean Segmentation Quality (mSQ) and mean Detection Quality (mDQ) over all classes (See notes in Appendix~\ref{sec:appendix-metrics-perf}). Model efficiency is evaluated with parameter number and latency.

\begin{figure}[!htb]
\captionsetup{justification=raggedright,singlelinecheck=false}
  \begin{minipage}{0.48\textwidth}
    \includegraphics[width=\linewidth]{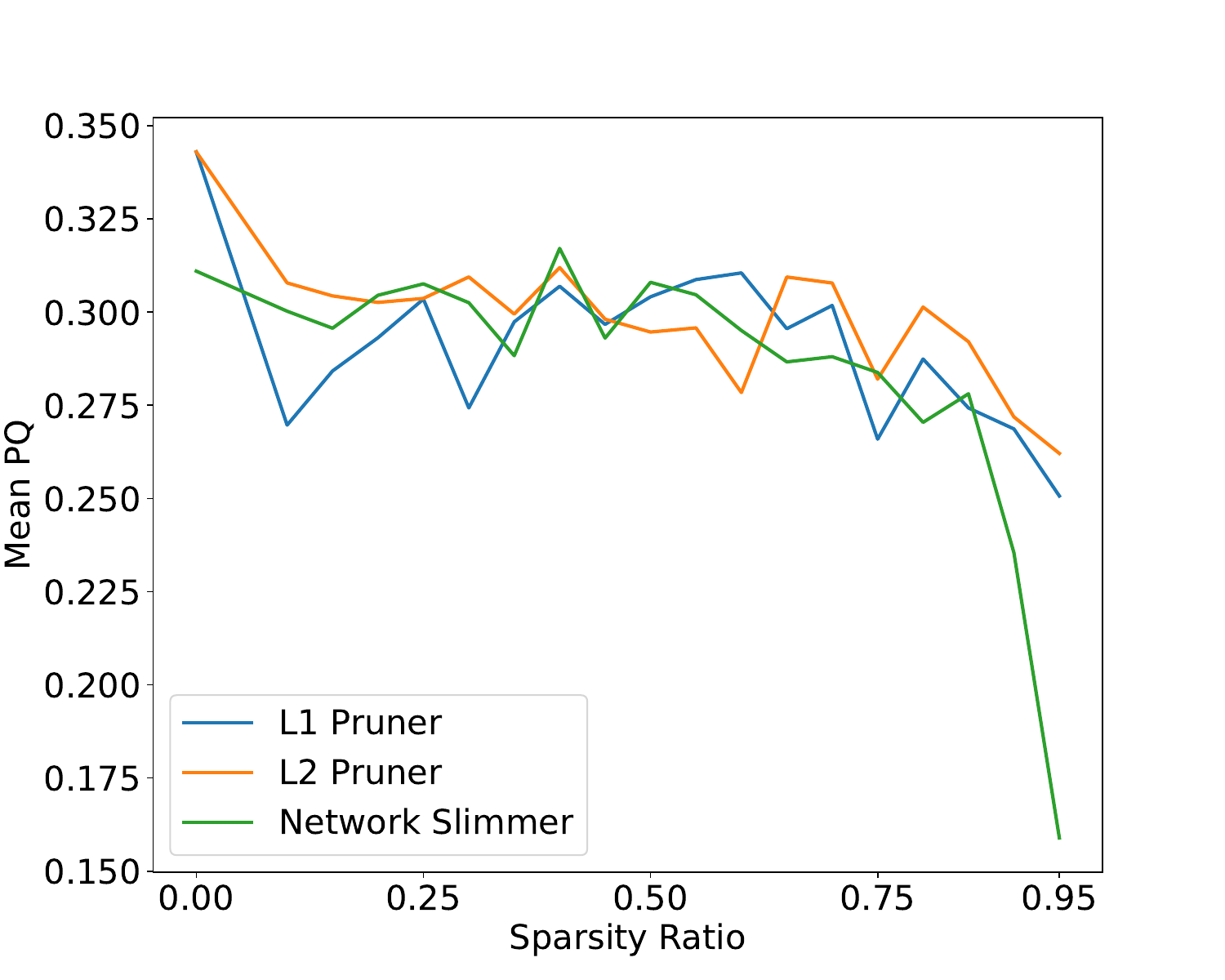}
  \end{minipage}
  \hfill
  \begin{minipage}{0.49\textwidth}
    \includegraphics[width=\linewidth]{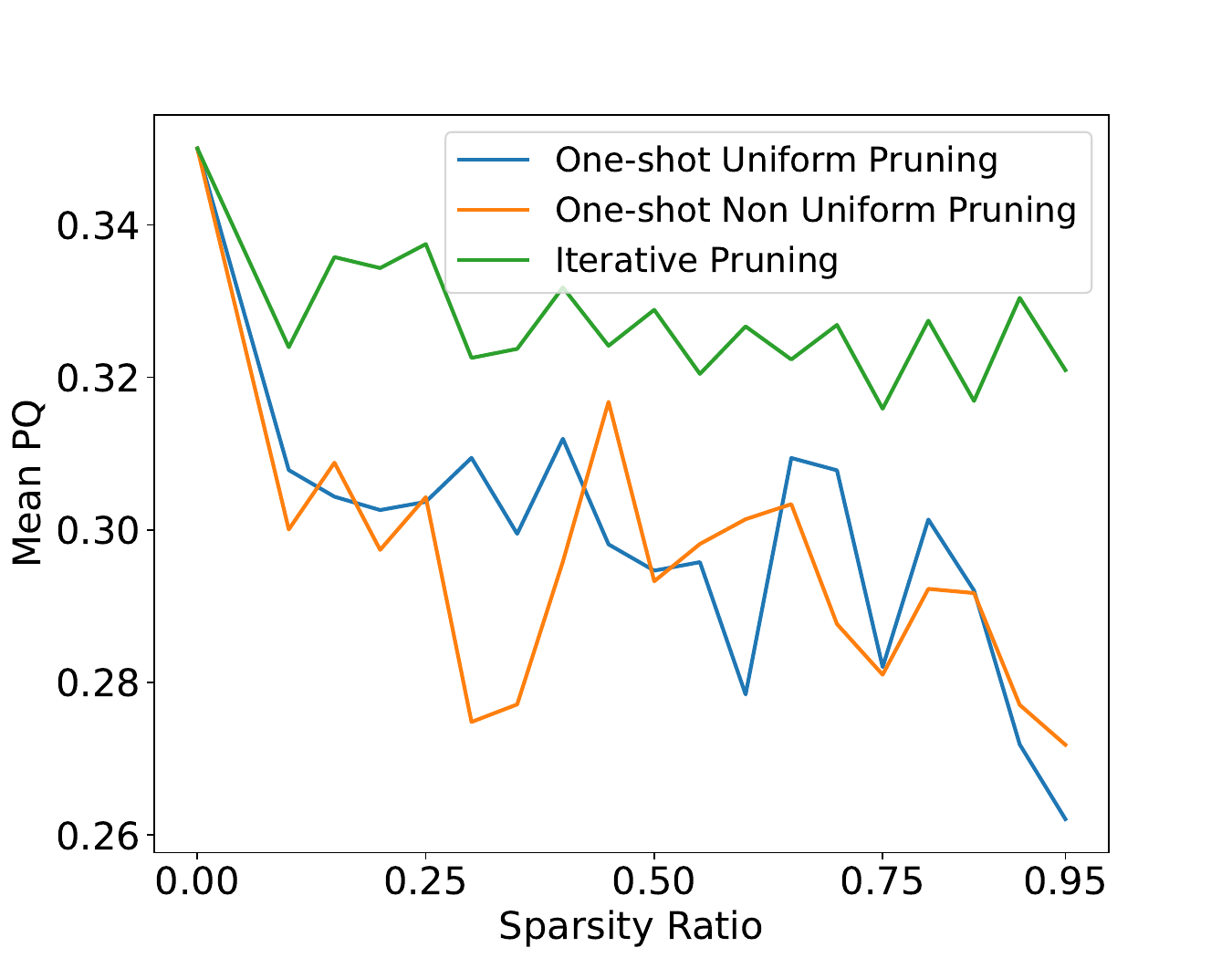}

  \end{minipage}
 
  \caption{The effects of filter importance heuristics (Left) and pruning strategies (Right) on mPQ.}\label{fig:pq-latency-combined}
\end{figure}

\noindent\textbf{One-shot Uniform Pruning.} One-shot uniform pruning followed by fine-tuning (\autoref{fig:pq-latency-combined} Left) led to a initial decrease of PQ by around 0.02 for L1/L2-norm pruners. However, pruning to higher sparsity levels did not lead to performance crashes with L1-norm and L2-norm pruners. Surprisingly, pruning up to 70\% sparsity level with L2-norm pruner achieved performance similar with that of 10\% sparsity level. On the other hand, model performance with network slimmer degraded drastically beyond 80\% sparsity level. See Appendix~\ref{sec:appendix-3folds-hovernetprune}  for all results from 3-fold testing.

\noindent\textbf{One-shot Non-uniform Pruning.} Intuitively, pruning performance might be improved by limiting the maximum sparsity level of interconnected layers with widespread impact across residual blocks and encoder-decoder layer. However, we did not observe superior model performance with such a strategy compared to one-shot uniform pruning. In addition, such a maximum sparsity level constraint limited the total reduction of model parameters from the interconnected layers and thus a smaller reduction in latency. See Appendix~\ref{sec:appendix-metrics-eff} for notes on the evaluation and efficiency metrics.

\noindent\textbf{Iterative Pruning vs One-shot Pruning.} Iterative pruning of 5\% filters with L2-norm heuristic consistently outperformed one-shot pruning (\autoref{fig:pq-latency-combined}) in terms of maintaining model performance. Surprisingly, pruning up to 85\% - 90\% sparsity levels resulted in not only identical model performance compared with pruning of low sparsity levels like 5\%, but also very similar performance as that of the baseline model. Critically, latency of the model reduced by 80\% (5-6 times faster) and memory footprint reduced by one magnitude, as reflected by the drastically reduced parameter numbers (\autoref{tab:hovernet_iterative}; see Appendix~\ref{sec:appendix-efficiency} for notes on efficiency improvement).
Due to the large number of experiments (5 strategies/heuristics $\times$ 19 runs $\times$ 3 folds), we kept the same hyperparameters, but one may further optimize performance by finer-grained hyperparameter tuning.

\begin{table}
\centering
\caption{Performance of iterative pruning of HoverNet (averaged over 3-fold).}
\label{tab:hovernet_iterative}
\scalebox{0.75}{
\begin{tblr}{
  column{even} = {c},
  column{3} = {c},
  column{5} = {c},
  column{7} = {c},
  hline{1-2,9} = {-}{0.08em},
}
\textbf{Sparsity} & \textbf{mPQ}    & \textbf{mSQ}    & \textbf{mDQ}     & \textbf{Latency} (ms) & \textbf{FLOPs} ($10^{10}$) & \textbf{Params} ($10^7$) \\
0.05              & 0.3239          & 0.8075          & 0.3968           & 676.76                & 27.57                    & 3.47                       \\
0.15              & 0.3343          & 0.8089          & 0.4079           & 638.06                & 23.41                    & 2.92                       \\
0.25              & 0.3225          & 0.7999          & 0.3953           & 540.09                & 19.61                    & 2.41                       \\
0.50              & 0.3204          & 0.8094          & 0.3915           & 225.85                & 11.55                    & 1.37                       \\
0.75              & 0.3274          & 0.8122          & 0.3986           & 148.90                & 5.56                     & 0.63                       \\
0.85              & 0.3304          & 0.8074          & 0.4024           & 120.65                & 3.74                     & 0.42                       \\
\textbf{0.90}     & 0.3209          & 0.8027          & 0.3934           & \textbf{105.43}       & \textbf{2.95}            & \textbf{0.33}              
\end{tblr}}
\end{table}




\subsection{Colorectal Cancer (CRC) Tissue Classification} 
We further assess model pruning in a second vision task, tissue classification, to investigate (1) whether a smaller network can be effectively pruned, (2) performance with different filter importance heuristics and (3) the impact of sparsity increments in iterative pruning on performance.

\noindent\textbf{Dataset and Implementation.} We leveraged a CRC dataset 
~\cite{kather2019predicting}, which contains 100,000 stain-normalized
224x224 patches at 20x sampled from H\&E whole-slide images (136 patients) with patch-wise labels of 9 tissue types. To establish a class-balanced dataset, we randomly selected 8700 image patches from each classes for training (7000 train; 2700 validation) and ran testing on the entire test set (7150 patches). We first trained ResNet18 ~\cite{He2015-ko} with AdamW optimizer at a learning rate of $5\times10^{-6}$ with weight decay of $1\times10^{-5}$ for at most 50 epochs with early stopping when validation loss did not improve for 5 consecutive epochs. We applied one-shot and iterative pruning at two different sparsity level increments (0.0625 and 0.25) followed by fine-tuning of at most 30 epochs with the same early stopping criteria. 

\noindent\textbf{Pruning Tssue Classification Model.} Similar to the much more complex HoverNet, ResNet18 can be effectively pruned: at 0.75 sparsity level, performance scores only dropped $<$3\% in one-shot pruning \autoref{tab:crc-heuristics} and with L2-norm iterative pruning \autoref{tab:crc}, while latency reduced to 1/3 and the parameter number reduced to 1/14 \autoref{tab:crc}. With one-shot pruning, L1-norm heuristic achieved slightly better performance than L2-norm and about 2\% better scores than Network Slimmer. Iterative pruning achieved about 1\% better performance scores than one-shot pruning. Iterative pruning with sparsity increments of 0.0625 and 0.25 achieved similar performance, suggesting the pruning performance is not very sensitive to this hyperparameter.
ResNet18 is much more prunable in our case than reported for natural scene image classification models~\cite{Li2016-vt}. such a difference potentially may be due to the different levels of information redundancy of these datasets.


\begin{table}
\centering
\caption{Performance of one-shot uniform pruning of ResNet18 CRC tissue classification model.}
\scalebox{0.8}{
\begin{tblr}{
  cell{1}{2} = {c=2}{},
  cell{1}{4} = {c=2}{},
  cell{1}{6} = {c=2}{},
  hline{1,3,7} = {-}{},
}
\textbf{\textbf{Sparsity}} & \textbf{L1 Pruner} &    & \textbf{L2 Pruner} &                                            & \textbf{Network Slimmer} &               \\
                           & Accuracy                & Weighted F1   & Accuracy                & Weighted F1            & Accuracy                 & Weighted F1   \\
0                          & 0.954$\pm$0.005           & 0.953$\pm$0.006 & 0.954$\pm$0.005           & 0.953$\pm$0.006          & 0.951$\pm$0.003            & 0.949$\pm$0.004 \\
0.25                       & 0.956$\pm$0.001           & 0.955$\pm$0.002 & 0.959$\pm$0.004           & 0.958$\pm$0.004          & 0.955$\pm$0.004            & 0.954$\pm$0.004 \\
0.50                       & 0.960$\pm$0.003           & 0.959$\pm$0.003 & 0.946$\pm$0.004           & 0.944$\pm$0.005          & 0.939$\pm$0.003            & 0.939$\pm$0.004 \\
0.75                       & \textbf{0.943$\pm$0.003}  & \textbf{0.942$\pm$0.003} & 0.937$\pm$0.003  & 0.936$\pm$0.002          & 0.924$\pm$0.008            & 0.921$\pm$0.009 
\end{tblr}}
\label{tab:crc-heuristics}
\end{table}

\begin{table}
\centering
\caption{Compare pruning performance for CRC tissue classification in one shot and iterative pruning with different sparsity increments with L2-norm pruners (averaged over 3 runs).} 
\scalebox{0.73}{
\begin{tblr}{
  cell{1}{2} = {c=2}{},
  cell{1}{4} = {c=2}{},
  cell{1}{6} = {c=2}{},
  cell{1}{8} = {c=2}{},
  hline{1,3,7} = {-}{},
}
\textbf{\textbf{Sparsity}} & \textbf{One shot} &             & \textbf{\textbf{Iterative 0.25}} &             & \textbf{Iterative 0.0625} &             & \textbf{Efficiency} &                             \\
                           & Accuracy          & Weighted F1 & Accuracy                         & Weighted F1 & Accuracy                  & Weighted F1 & Latency (ms)        & Params ($10^{6}$) \\
0                          & 0.954             & 0.953       & 0.954                            & 0.953       & 0.954                     & 0.953       & 6.82                & 11.70                       \\
0.25                       & 0.959             & 0.958       & 0.959                            & 0.958       & 0.949                     & 0.948       & 5.65                & 6..67                       \\
0.50                       & 0.946             & 0.944       & 0.948                            & 0.946       & 0.944                     & 0.943       & 3.57                & 3.05                        \\
0.75                       & 0.937             & 0.936       & 0.948                            & 0.948       & 0.949                     & 0.948       & \textbf{2.17}       & \textbf{0.83}               
\end{tblr}}
\label{tab:crc}
\end{table}

\section{Conclusion}
In this study, we investigated model pruning in digital pathology applications. We proposed an effective pruning strategy to reduce the computation budget of HoverNet, a widely adopted architecture for nuclei instance segmentation and classification.
We further demonstrated effective model pruning in a much smaller model, ResNet18, for tumor classification. 
Our observations suggest that large models are not necessarily a hard requirement for reliable and effective inference in digital and computational pathology applications.
The pruned models, compact and efficient, may enable the deployment of AI in resource-constrained clinical sites and onto edge devices, for example, enabling AI on whole-slide scanners. 
For future work, we plan to (1) assess the robustness of pruned models compared with original models \cite{holste2023does} (more discussion in Appendix~\ref{sec:appendix-robustness}) (2) combine pruning with quantization and deploy the pruned model on edge devices and (3) investigate pruning for Vision Transformers models like CellViT~\cite{hörst2023cellvit}. 

\bibliography{ref}

\newpage
\appendix
\renewcommand{\thetable}{{S\arabic{table}}}
\renewcommand{\thefigure}{S\arabic{figure}}
\setcounter{figure}{0}

\section{Notes on Knowledge Distillation}\label{sec:appendix-KD}

Knowledge Distillation (KD) is the process of transferring knowledge from a large model (teacher model) to a relatively smaller model (student model)~\cite{Hinton2015-wp}. KD has been shown to be effective in training a smaller model (e.g. ResNet-50) to the same level of performance as the larger teacher model (e.g. ResNet-152). However, the level of compression achieved by KD is relatively lower than some other techniques like pruning, because the selected student model is usually an existing well-designed architectures. KD has also been explored in digital pathology for model compression and to improve the performance of the student model. For e.g., Javed et al., proposed a knowledge distillation-based method to improve the
performance of shallow networks for tissue phenotyping in
histology images~\cite{javed2023knowledge}. They used ResNet-18 as a student model and Resnet-50 as a teacher model. The student model can be further pruned or quantized depending on the need~\cite{9523139}. DiPalma et al. proposed a method 
for improving the computational efficiency of histology image classification~\cite{dipalma2021resolutionbased}. The authors proposed to train the teacher network on the high-resolution images and the student network on the low-resolution images. 

It is important to note that KD as a generic concept/method has find its way in many other applications than model compression, for example, for the following two studies on self-supervised learning, the teacher and student models are of the same architecture. Yu et al.\ proposed SLPD, which encoded the intra- and inter-slide semantic structures by modelling the mutual-region/slide relations using knowledge distillation~\cite{yu2023slpd}; Luo et al.\ proposed a negative instance-guided, self-distillation framework to directly train an instance-level classifier end-to-end as an alternative to Multiple Instance Learning methods~\cite{10195157}. 

\section{Semi-Structured Pruning}\label{sec:appendix-semistruct}
In contrast to structured pruning, where an entire filter is completely removed, 
semi-structured pruning explores sparsity patterns between unstructured and 
structured pruning, such as block sparsity or n:m sparsity. Unlike unstructured 
pruning, some of the new hardware (NVIDIA A100) can leverage n:m sparsity to speed up inference.

\section{Efficiency Metrics and Performance Metrics}
\label{sec:appendix-metrics}
\subsection{Performance Metrics} \label{sec:appendix-metrics-perf}
Following \cite{Gamper2020-rc}, we used the metrics (PQ, DQ, SQ) detailed below for evaluating the performance of the pruned model.

We report mean PQ, DQ and SQ over all classes by first pooling all positive instances from all test images and then calculate these metrics for each class, followed by averaging over classes. Note that in the PanNuke dataset paper ~\cite{Gamper2020-rc}, the authors calculated mean PQ by first calculating PQ for each class from each image and then averaging over the number of images for these images with positive instances for corresponding classes. We found that PQ and DQ are similar with these two ways of calculation, SQ are about 0.2 higher in our approach.

\begin{enumerate}


    \item Detection Quality (DQ): DQ is a widely used metric for evaluating segmentation models. DQ is the F1 score that measures the quality of instance detection (Graham et al. 2018). Mathematically, it can be described as:

        \begin{equation*}
            DQ = \frac{|TP|}{|TP| + \frac{1}{2}|FP| + \frac{1}{2}|FN|},
        \end{equation*}
        where TP, FP and FN denote True Positive, False Positive and False Negative respectively.
    
    \item Segmentation Quality (SQ): SQ measures how close each correctly detected instance is to its ground truth mask. Mathematically, it can be described as:
    \begin{equation*}
            SQ = \frac{\Sigma_{(x,y)\in TP} IoU(x,y)}{|TP|},
    \end{equation*}
    where IoU is the intersection over the union. We only considered instances with IoU $\geq$ 0.5, which is proven to have unique matches between ground-truthc and predicted instances (Kirillov et al. 2018).

     \item Panoptic Quality (PQ): PQ, introduced by Krillov et al. (Kirillov et al. 2018), is a unified score for comparing both segmentation and classification between the predicted labels and ground truth. Mathematically, PQ is the product of DQ and SQ:

    \begin{equation*}
                    PQ = \frac{|TP|}{|TP| + \frac{1}{2}|FP| + \frac{1}{2}|FN|} \times \frac{\Sigma_{(x,y)\in TP} IoU(x,y)}{|TP|}
    \end{equation*}

\end{enumerate}

\section{Intra-residual block and inter-residual block pruning}
\label{sec:appendix-prune-resnet}
Example illustration of intra-residual block (\figureref{fig:intrablock}) and inter-residual block pruning (\figureref{fig:interblock}).
\begin{figure}[htbp]
    \floatconts
    {fig:intrablock}
    {\caption{Example intra-block pruning of ResNet18 up to the 3rd residual block. In each residual block, the output dimension (channels) of the 1st convolution layer, the subsequent batch normalization layer as well as the input dimension (channels) of the 2nd convolution layer are pruned with the same ratio and indexing, determined by, for example, the filter importance ranking with L1/L2 pruners. Grayed channel dimensions are pruned with the pruning ratio of 1/a, 1/b, 1/c, etc.}}
    {\includegraphics[width=.95\linewidth]{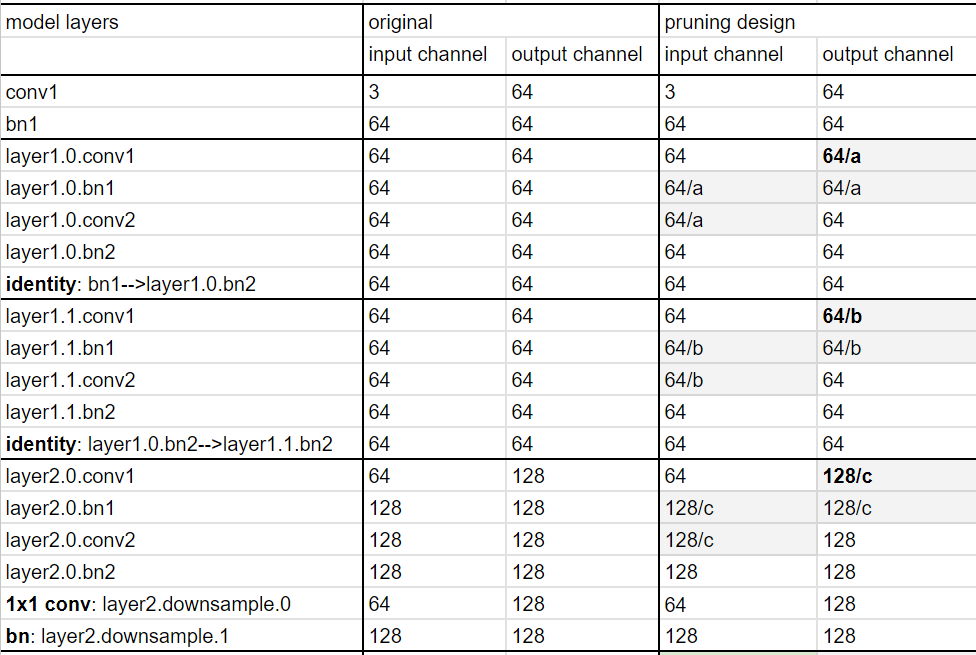}}  
\end{figure}
\begin{figure}[htbp!]
\floatconts
    {fig:interblock}
    {\caption{Example inter-block pruning of ResNet18 up to the 3rd residual block. The yellow highlighted channel dimensions belong to a inter-connected group of layers, and thus the same pruning rario (1/i) and pruning indexing should be applied to match the channel dimensions after pruning. The bold blue highlighted channel dimension (e.g. conv1 output channel dimension) were used in our study for ranking the filter importance in L1/L2 pruners. The green highlighted channel dimensions belong to another interconnected group, only part of which is shown in this illustration. Note that 1x1 conv for downsampling enabled the green highlighted group of layers to become an independent group. Intra-channel pruning is also illustrated with the grayed channel dimensions along with their pruning ratios.}}
    {\includegraphics[width=.95\linewidth]{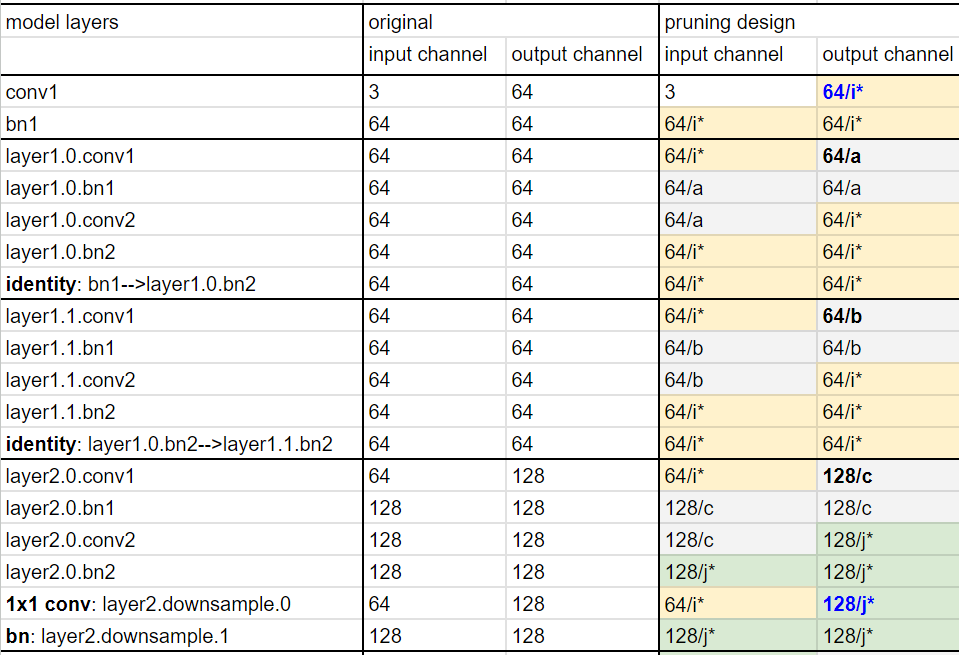}}  
\end{figure}

\FloatBarrier

\subsection{Efficiency Metrics} \label{sec:appendix-metrics-eff}
Most existing literature adopts theoretical floating point operation (FLOPs) as their metric for assessing the speed-up or computation saving after pruning. FLOPs refers to the theoretical number of arithmetic operations required to perform a given computation. FLOPs are generally proportional to the model size, though dependent on what exact layers designed in a particular model.  

However, theoretical FLOPs does not directly reflect model speedup nor memory saving and thus at most a supplementary metric when evaluating model efficiency changes before and after model pruning/compression \cite{ma2018shufflenet}. One main reason is that for different types of layers and even for the same type of layers with different design choices, the same number of floating number operation reduction does not equal the same time/memory saving. One example is a convolution layer with filter size of 3x3 versus 5x5. Most modern scientific computing libraries include optimized implementation for 3x3 but not 5x5 convolution operations and thus 3x3 is faster than 5x5. Another example is that plain convolution networks like VGGs, which do not include residual blocks, generally are more efficient than residual networks, and as such even when removing the same number of parameters, the former will accelerate more than the latter. See \cite{ma2018shufflenet} and \cite{fu2022depthshrinker} for more in-depth assessment and explanations.

We thus turn to the following two complementary metrics for our assessment.
\textbf{Latency}:  Latency refers to the time taken for a model to process input or a batch input, where input refers to images with certain height, width and channels (like R,G,B channels). To get accuracy latency, we (1) fix the GPU device type when comparing different experiment settings (2) fix the image batch size and image dimensions (3) perform GPU warm-up before the actual latency computation for our models and (4) since GPU computing is asynchronous (different threads execute computation at slightly different timing), we wait till all computing threads stop before ending the timing for latency calculation.
\textbf{Model parameter number}: The number of model parameters along with the data type are directly related to the model size. For example, one can calculate that a model with 1 million parameters whose weight data type of float32 roughly occupies 4 megabyte of disk/dynamic memory.

Throughput refers to the number of input images processed by a model within a given time. Despite that multi-processing in general can increase latency and thus may not directly reflect model efficiency changes, one can fix the number of GPUs used for throughput calculation and use this metric along with latency, both are similar metrics that measures time-related improvement of model pruning/compression.

\section{Model performance across folds for various pruning heuristics and sparsity ratios}
\label{sec:appendix-3folds-hovernetprune}
\figureref{fig:pqfolds} shows detailed performance of model from each of the 3 folds across sparsity ratios of 0.1 to 0.95 at 0.05 increments for various pruning heuristics. 
\begin{figure}[htbp]
\floatconts
    {fig:pqfolds}
    {\caption{Per fold PQ for various pruning heuristics and sparsity ratios}}
    {\includegraphics[width=.95\linewidth]{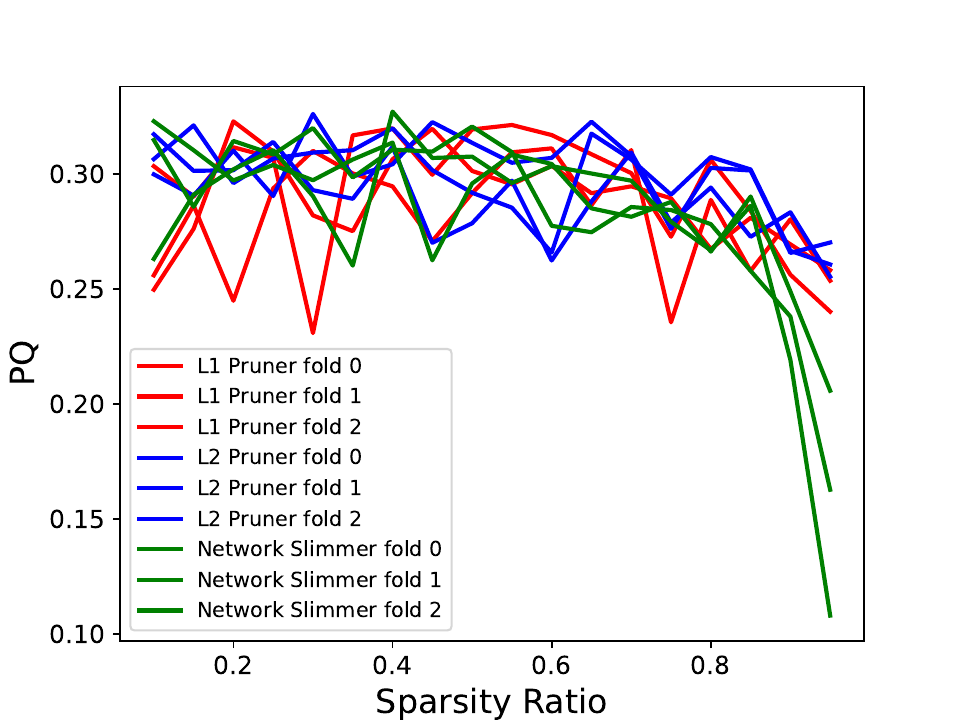}}
\end{figure}

\section{Notes on reduction of latency and model parameter numbers in model pruning}
\label{sec:appendix-efficiency}
\textbf{Reduction of model parameter number} It is obvious that the model parameter number does not linearly reduce with the sparsity levels, including in \autoref{tab:crc}. Let a convolution layer has $i$ input channels, $o$ output filters, kernel height and width of $k$, its number of parameters is $i\times o \times k \times k $. In our pruning setting, we pruned all layers. For this convolution layer, when the input and output dimensions are both pruned with, for example, a sparsity level of 0.25, the remaining input and output channel numbers become $0.75\times i$ and $0.75\times o$, resulting in $ 0.75 \times i\times 0..75\times o \times k \times k$ number of parameters after pruning. As such, pruning of this convolution layer reduced more than 25\% of its parameters. Since most of the parameters in a ResNets are in the convolution layers, it is not surprising that model parameter reduction is faster than linear against sparsity levels during pruning.

\textbf{Reduction of latency} In \autoref{tab:crc}, model latency reduced to 1/3 of original model, disproportional to the reduction of model parameters (1/10). First, such a behaviour is common in literature, such as for the original works of L1/L2 norm and network slimmer. Second, despite that filter pruning removed large proportion of filters, the residual connections were kept, which is know to be less efficient than plain convolution layers \cite{fu2022depthshrinker}. Third, the exact reduction of latency for each particular layers is highly related to the implementation of the model training framework (PyTorch) and hardware level implementation (e.g. CuDNN). Fourth, imbalanced convolution channel input/output (
convolution layers with input and output channel sizes differ a lot) was shown to contribute to slowing down model inference speed \cite{ma2018shufflenet}. Resnets with some of these layers may not be as readily speed up even after pruning.


\section{Discussion on Model Robustness}\label{sec:appendix-robustness}
Healthcare is a sensitive domain and therefore, it is important to discuss the effect of pruning on model bias and robustness. Hooker et al.,  studied model compression in detail and observed that both pruning and compression exacerbate the algorithmic bias~\cite{hooker2020characterising}. Hooker et al., empirically showed that while the overall accuracy remains unchanged but some classes bear a disproportionately high portion of the error. There has also been work in the literature to preserve adversarial robustness during pruning~\cite{jian2022pruning}.

\end{document}